\documentclass[epj]{svjour}
\usepackage{graphicx}
\usepackage{amsmath}
\usepackage{amsfonts}
\usepackage{amssymb}

\newcommand{\be}{\begin{equation}}
\newcommand{\ee}{\end{equation}}

\begin{document}
\title{Statistical properties of power-law random banded unitary matrices in the
delocalization-localization transition regime}
\author{Jayendra N. Bandyopadhyay\inst{1}\inst{,}\thanks{\emph{Present address: Department of Physics, Birla Institute of Technology and Science, Pilani 333031, India}} 
\and Jiangbin Gong\inst{1}\inst{,}\inst{2}}                     
%!TEX encoding = UTF-8 Unicode
\institute{Department of Physics and Centre for
Computational Science and Engineering, National
University of Singapore, 117542, Republic of
Singapore \and 
NUS Graduate School for Integrative
Sciences and Engineering, 117597 Singapore,
Republic of Singapore}
\abstract{
Power-law random banded unitary matrices (PRBUM), whose matrix
elements decay in a power-law fashion, were recently proposed to
model the critical statistics of the Floquet eigenstates of
periodically driven quantum systems. In this work, we numerically study in
detail the statistical properties of PRBUM ensembles in the
delocalization-localization transition regime. In particular,
implications of the delocalization-localization transition  for
the fractal dimension of the eigenvectors, for the distribution
function of the eigenvector components, and for the nearest neighbor
 spacing statistics of the eigenphases are examined.  On the one
hand, our results further indicate that a PRBUM ensemble can serve
as a unitary analog of the power-law random Hermitian matrix model
for Anderson transition. On the other hand, some statistical features unseen before
are found from PRBUM. For example, the dependence of the fractal dimension of the eigenvectors of PRBUM
upon one ensemble parameter displays features that are quite
different from that for the power-law random Hermitian matrix model.
Furthermore, in the time-reversal symmetric case the nearest neighbor spacing
distribution of PRBUM eigenphases is found to obey a semi-Poisson distribution
for a broad range, but display an anomalous level
repulsion in the absence of time-reversal symmetry.
} 

\maketitle

\section{Introduction}

As a statistical description with few parameters, random matrix
theory (RMT) \cite{mehta} has found important applications in a wide
variety of topics, such as complex heavy nuclei \cite{porter},
disordered electronic systems \cite{disorder}, large complex atoms
\cite{camrada}, stock market data analysis \cite{market},
atmospheric data analysis \cite{santh}, analysis of human
electroencephalogram (EEG) \cite{eeg}, and spectra of complex
networks \cite{networks}, to name a few. Many interesting extensions
of RMT have also been proposed to model different aspects of complex
systems \cite{shukla}. More relevant to this study, we note that
RMT has been exploited extensively to predict the statistics of
quantum systems with chaotic classical limits
\cite{haakebook,reichlbook}. In these kind of systems, it is now almost 
confirmed that the nearest neighbor spacing
distribution (NNSD) of energy levels follow the standard RMT
prediction, namely, the Wigner-Dyson statistics \cite{bohigas}.
On the other hand, quantum systems with regular classical limit, 
the same NNSD follows the Poisson statistics \cite{berry-tabor}.

The (Anderson) delocalization-localization transition
\cite{anderson,AT} in disordered systems is a topic of enormous
interest in condensed-matter physics because it is a metal-insulator
transition induced by disorder. Recent advances in cold-atom
physics have also made it possible to directly observe this
transition under controlled disorder \cite{exp1,exp2}. For weak
disorder strength the eigenstates of the disordered system are
essentially structureless and delocalized rather homogeneously over
the whole Hilbert space. As a result the eigenstates can overlap
very well with each other and level repulsion as a characteristic of
the Wigner-Dyson statistics is anticipated. For strong disorder
strength, the eigenstates are highly localized and their overlap is
negligible. The energy levels are hence uncorrelated and level
clustering as a feature of the Poisson statistics emerges.
Interestingly, in the vicinity of the delocalization-localization
transition, the eigenstates display multifractal features \cite{AT}
(thus neither delocalized nor localized) and the energy level
statistics shows an interesting hybrid of the Wigner-Dyson
statistics and the Poisson statistics \cite{shapiro}. To model such
intermediate statistics for disordered systems undergoing the
delocalization-localization transition, the conventional RMT must be
extended. Indeed, the power-law random banded matrix (PRBM) ensemble
was proposed with success \cite{AT,prbm-pap}. A PRBM ensemble has
two more parameters than a conventional random matrix, and the
matrix elements of a PRBM decay as a power-law function of the
distance from the diagonal.

The delocalization-localization transition and the associated
critical statistics may also occur in a quantum system in the
absence of any disorder, e.g., in a two-dimensional electron gas
subject to a perpendicular magnetic field \cite{harper,holthaus} or
in driven quantum systems
\cite{casati,jiao,exp2,kicked-harper,JJ1,JJ2}. For these cases,  the
well-established connection between classical chaos and the
conventional RMT predictions, or between classical integrability and
the Poisson statistics, breaks down. Of particular interest here is
the statistics of the Floquet states of periodically driven systems.
Indeed,  critical eigenstate statistics and critical spectral
statistics have been found in the kicked-Harper model
\cite{kicked-harper}, in an on-resonance double-kicked rotor model
\cite{JJ1}, and most recently, in a double-kicked top model
\cite{JJ2}, regardless of whether the underlying classical limit is
regular or chaotic.  These findings motivated us to propose in Ref.
\cite{JJJ} an extension of Dyson's standard random unitary matrix
ensemble \cite{dyson}, such that the delocalization-localization
transition in quantum mapping systems can be modeled with few
parameters. Somewhat analogous to the Hermitian PRBM, we proposed
in Ref. \cite{JJJ} a ``power-law random banded {\it unitary}
matrices (PRBUM)" ensemble, with the elements of each unitary matrix
decaying as a power-law function of their distance from the
diagonal. Remarkably, as shown in Ref. \cite{JJJ} in terms of a
fractal dimension analysis of the Floquet eigenstates, PRBUM around the
localization-delocalization transition regime
agrees with the statistics of an actual
driven quantum system in the presence or absence of time-reversal
symmetry, whereas the PRBM does not agree.  This indicates that
PRBUM is not a trivial extension of PRBM from Hermitian matrices to
unitary matrices. This being the case, it becomes necessary to
investigate from many different angles the statistical properties of
PRBUM. More
specifically, we examine in this work the implications of the
delocalization-localization transition for the fractal dimension of
PRBUM eigenvectors, for the distribution function of PRBUM
eigenvector components, and for the NNSD of PRBUM eigenphases.
Interesting findings include the dependence of the fractal dimension
of PRBUM eigenvectors upon ensemble parameters, comparison between
PRBUM eigenvector component statistics with a normal distribution,
transition of the NNSD from the Wigner-Dyson statistics to the
Poisson statistics, and anomalous eigenphase level repulsion of
PRBUM without time-reversal symmetry. Besides these driven quantum 
critical systems, the PRBUM ensemble has already found one application 
in a very interesting and important field of research. This ensemble 
has been found to be relevant to understanding sound propagation 
through underwater \cite{Steve}. Our detailed results are hoped to motivate more interests in the potential applications of PRBUM. 

This paper is organized as follows. In Sec. \ref{sec2}, we briefly
introduce PRBUM first proposed in Ref. \cite{JJJ}. In Sec.
\ref{sec3}, we study the fractal dimension of PRBUM eigenvectors in
the delocalization-localization transition regime as a function of
two ensemble parameters. Due to the finite size of the system we can afford to
study computationally, it is not feasible to identify
the exact transition point. For this reason we examine the transition point from different
perspectives in Sec.~\ref{sec4}. Statistical properties of the eigenvector
components of PRBUM are examined in Sec. \ref{sec5}. In Sec.
\ref{sec6}, we seek the signature of the delocalization-localization
transition in the NNSD of PRBUM eigenphases. Finally, we conclude this paper in Sec. \ref{sec6}.

\section{From Dyson's circular ensembles to PRBUM ensembles}
\label{sec2}

Dyson introduced three types of random unitary matrix
(circular) ensembles to study statistical properties of complex quantum
systems \cite{dyson}. Dyson's circular ensembles have also proved to
be useful to study periodically driven quantum systems whose
classical limits are chaotic. In particular, the circular unitary
ensemble (CUE) models complex quantum systems without time-reversal
symmetry. The CUE is an ensemble of ($N \times N$) random unitary
matrices distributed with the natural Haar measure on the unitary
group $\mathbb{U}(N)$. The Haar measure is analogous to a uniform
distribution which implies that the probability of having any
unitary matrix in the CUE is equal. The eigenphases of CUE are hence
uniformly distributed on the unit circle.  The circular orthogonal
ensemble (COE) is an ensemble of symmetric unitary matrices that
model systems with time-reversal symmetry. A COE matrix $U$ can be
expressed in terms of another unitary matrix $W$ from CUE, i.e.,
$U=W^{T}W$. In addition, sampling a matrix $U$ from COE is
equivalent to sampling a matrix $W$ from CUE \cite{karol}.  The
third type of Dyson's circular ensemble is not related to this work
and hence will not be discussed here.

\begin{figure}[b]
\centering
\includegraphics[width=8.4cm,height=4cm]{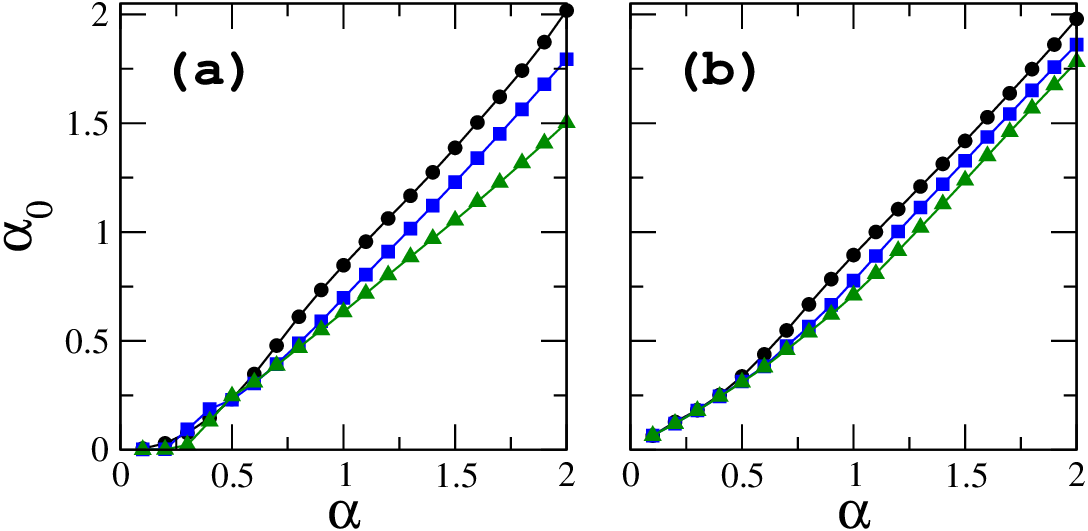}
\caption{(Color online) The parameter $\alpha_0$ is plotted as a function
of the parameter $\alpha$ for (a) PRBUM-COE and (b) PRBUM-CUE. Here
$b=0.1\, ({\bullet}), 0.3\, (\blacktriangle)$, and $0.5\, (\blacksquare)$.}
\label{alphVsalph0}
\end{figure}

One convenient numerical algorithm for the generation of CUE and COE
is Mezzadri's algorithm \cite{mezzadri}. The first step in this
algorithm is to sample matrices from an ensemble of $N \times N$
random matrices $\{Z_{G}\}$, with both the real and the
imaginary parts of the matrix elements being Gaussian distributed
random numbers (such a complex
matrix ensemble is known as the Ginibre ensemble in the literature).
As already detailed in Ref. \cite{JJJ}, the key element in our
algorithm for PRBUM generation is to replace $\{Z_{G}\}$ by matrices
drawn from the PRBM
ensemble, with all other following steps being still the same as in
the original Mezzadri's algorithm.  The PRBM ensemble is composed of
random Hermitian matrices whose matrix elements $\{H_{ij}\}$ are
sampled from independently distributed Gaussian random numbers with
the mean $\langle H_{ij}\rangle = 0$ and the variance
\be
\sigma^2(H_{ij}) = 2\left[ 1 + \left(\frac{|i-j|}{b}\right)^
{2\alpha}\right]^{-1}.
\label{var-PRBM}
\ee
Since PRBM has two
parameters $\alpha$ and $b$ (the case of $\alpha = 1.0$ represents
the critical Anderson transition point in a time-independent
problem), the PRBUM generated from PRBM can also be characterized by
$\alpha$ and $b$. Parallel to the CUE and COE matrices generated
from Mezzadri's algorithm, our modified algorithm generates CUE and
COE versions of PRBUM ensembles, which are called PRBUM-CUE and
PRBUM-COE below.

\begin{figure}[t]
\centering
\includegraphics[width=8.4cm,height=4cm]{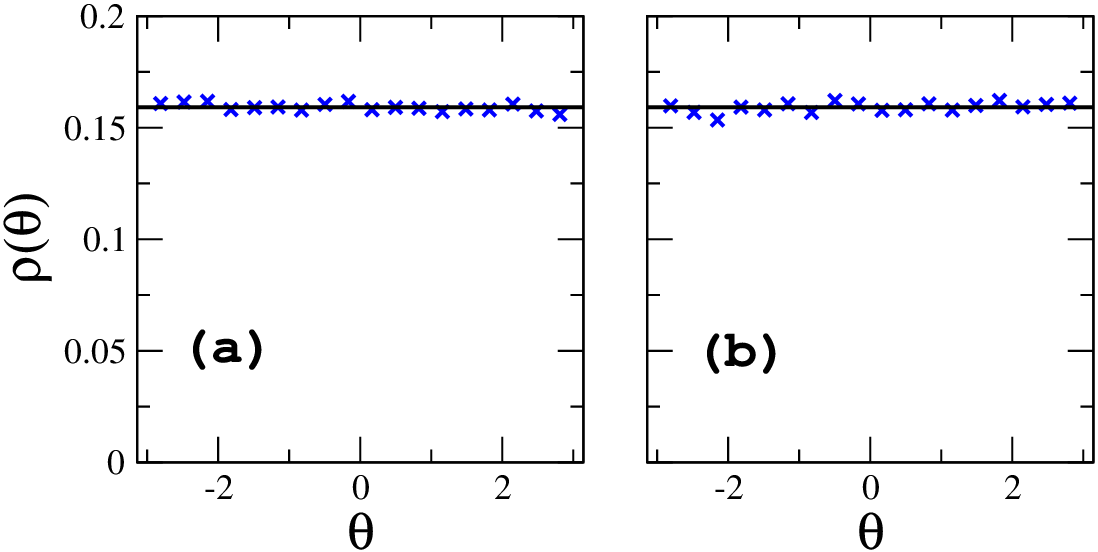}
\caption{(Color online) Normalized distribution [denoted
$\rho(\theta)$] of the eigenphases $\theta$ for (a) PRBUM-COE
and for (b) PRBUM-CUE. Here the parameters $b=0.1$, $\alpha =
0.5$. The matrix dimension $N=50$ and the ensemble size is 10000.
Numerical results here indicate a uniform eigenphase density.}
\label{qe-density}
\end{figure}

Other details about the generation of PRBUM can be found in Ref.
\cite{JJJ}. Remarkably, the variance of the matrix elements of the generated unitary
matrices does display a power-law feature, i.e.,
\be
\sigma^2(U_{ij}) = a_0 \left[ 1 + \left(\frac{|i-j|}{b_0}
\right)^{2\alpha_0}\right]^{-1}.
\label{var-PRBUM}
\ee
In Fig. \ref{alphVsalph0}, we present the parameter $\alpha_0$
of the PRBUM ensemble as a function of the parameter $\alpha$ of the initial
PRBM ensemble. This result shows that the numerically found values of $\alpha_0$
depend on the parameters $\alpha$ and $b$ of the
PRBM used in the first step of our algorithm. Moreover, even
with the same set of $\alpha$ and $b$, the values of $\alpha_0$
for PRBUM-CUE differ from those for PRBUM-COE. For this
reason, we still prefer to use the parameters $\alpha$ and $b$ to
characterize our PRBUM. Figure \ref{qe-density} depicts the
normalized density distribution of the eigenphases for both
PRBUM-COE and PRBUM-CUE. Eigenphase $\theta$ of a unitary operator
$U$ is defined in the following way:
\be
U |\phi_\lambda\rangle = \exp(i \theta_\lambda) |\phi_\lambda\rangle,
\ee
where $\{|\phi_\lambda\rangle\}$ is an eigenstate of $U$.
The distribution is found to be uniform as
in the CUE and COE cases. Here we show result only for $b=0.1$ with
$\alpha = 0.5$. For other values of $\alpha$, we also find similar
uniform density. This result is important because it indicates
that the PRBUM ensemble also satisfies Haar measure and our algorithm
is very robust. As such, there is no
need to carry out an unfolding procedure when we study the NNSD.

\section{Fractal dimension analysis of PRBUM eigenvectors}
\label{sec3}

The fractal statistics of PRBUM eigenvectors can be analyzed by the
inverse participation ratio (IPR),
\be P_2^{(\lambda)} = \sum_m
|\langle m | \phi_\lambda\rangle|^4,
\ee
where $\lambda$ is the
index for eigenstates $| \phi_\lambda\rangle$ and $\{|m\rangle\}$
are generic basis states. In our early study \cite{JJJ}, it was
found that for $\alpha=1.0$, the IPR scales anomalously (i.e., $D_2$ is 
fraction) with the Hilbert space dimension $N$, i.e.,
\be
P_2^{(\lambda)} \sim N^{-D_2^{(\lambda)}},
\label{calc_D2}
\ee
where $D_2^{(\lambda)}$ is
just the fractal dimension of a particular eigenstate
$|\phi_\lambda\rangle$. Following the methodology in studies of the
PRBM ensemble \cite{mirlin1,mirlin2}, it was further observed in
Ref. \cite{JJJ} that for sufficiently large $N$, the distribution of
$\ln (P_2^{(\lambda)})$ for $\alpha=1.0$ displays evidence of being 
scale-invariant. This scale-invariance, which also suggests self-similarity,
is valid upto a certain scale. We define $D_2$ in the non-critical
regime where, at the thermodynamic limit, the system should be either in 
delocalized ($D_2 = 1.0$) or localized ($D_2 = 0.0$) regime. However,
due to the bounded scale, some part of the delocalized or localized regime 
(close to the critical point) may show self-similarity like at the critical point
with different fractional value of $D_2$.

Basically the technique of determining $D_2$ is following: due to the 
scale-invariance, the distribution function of $\ln
(P_2^{(\lambda)})$ only shifts as $N$ varies. As such the spectral
average of $\ln (P_2^{(\lambda)})$, denoted by $\langle \ln (P_2)
\rangle$, is linearly related to $\ln N$. The slope of the
$\langle \ln (P_2)\rangle $ vs $\ln N$ curve gives a single
fractal dimension $D_2$ for the system. The scaling of the variance
of $P_2^{(\lambda)}$ with $N$ is also studied in Ref. \cite{JJJ} for
$\alpha=1.0$, with features different from those observed in PRBM. We
have also observed that conclusions drawn in Ref. \cite{JJJ} based
on PRBUM for $\alpha=1.0$ are also valid if we apply our analysis to a
different random unitary matrix model that also aims to describe
fractal and intermediate statistics in quantum mapping
systems~\cite{french,jayendraun}.

\begin{figure}[b]
\centering
\includegraphics[width=8.4cm,height=4cm]{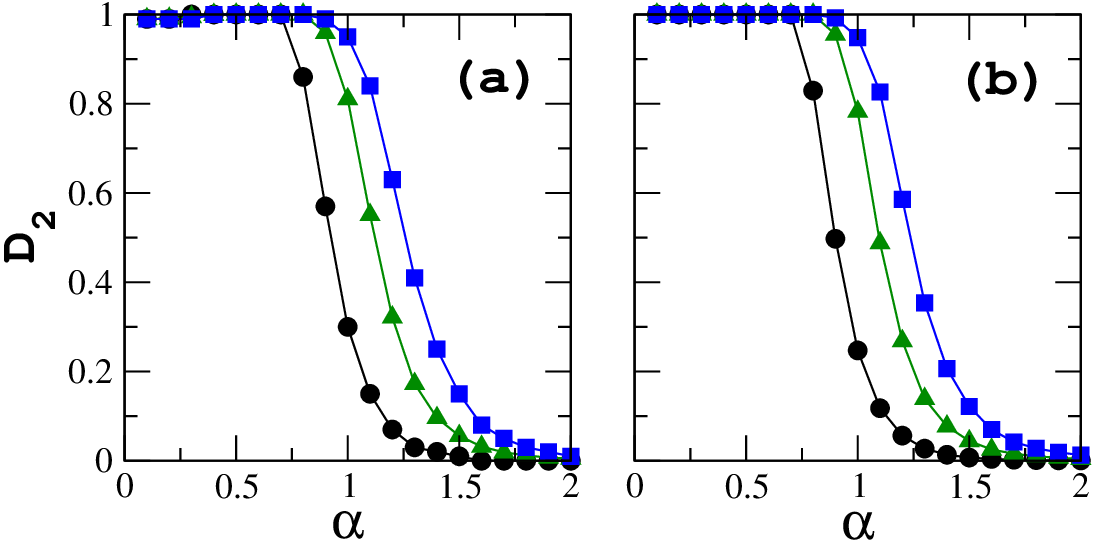}
\caption{(Color online) The fractal dimension $D_2$ of the eigenvectors
of PRBUM plotted as a function of $\alpha$, for $b=0.1 ({\bullet}),
0.3 (\blacktriangle)$, and $0.5 (\blacksquare)$.
For both (a) PRBUM-COE and (b) PRBUM-CUE, one may identify that the
delocalization-localization transition occurs around $\alpha = 1.0$. }
\label{alphVsD2}
\end{figure}

The purpose of this section is twofold. First, we present further
numerical evidence that like PRBM, the case of $\alpha=1.0$ is at least
(if not exactly) close to the critical transition point of PRBUM.
We do so by examining the transition of the fractal dimension $D_2$ as a function of
$\alpha$ for various values of $b$. Second, we examine the dependence
of the fractal dimension $D_2$ upon $b$.  Note that since we are not
in the thermodynamic limit, so the $D_2$ values only represent some
self-similar properties within a certain scale.

In Fig.~\ref{alphVsD2}, we show the dependence of the numerically
obtained $D_2$ as a function of $\alpha$ for three different values
of $b$, i.e., $b = 0.1, 0.3$ and $0.5$.
Similar to Ref. \cite{JJJ}, we have considered matrix sizes ranging from
a few hundreds to a few thousands to extract the value of $D_2$.
Left and right panels of Fig.~\ref{alphVsD2} are
for PRBUM-COE and PRBUM-CUE, respectively. For both cases, as the
value of $\alpha$ scans through unity, we have the change from $D_2
\sim 1.0$ to $D_2 \sim 0.0$, which reflects the transition from
extended states to localized states. This is always the case despite
the big change in the values $b$ (see discussion below). In this sense,
our PRBUM ensemble may be regarded as another unitary model of the
Anderson transition \cite{AT_unitary}. However, the
delocalization-localization transition is not very sharp; eigenstates are
fractal ($1.0 > D_2 > 0.0$) for a somewhat wide range of values of $\alpha$ around
$\alpha = 1.0$. We will discuss more on this issue in the next section.
\begin{figure}[t]
\centering
\includegraphics[width=8.4cm,height=7cm]{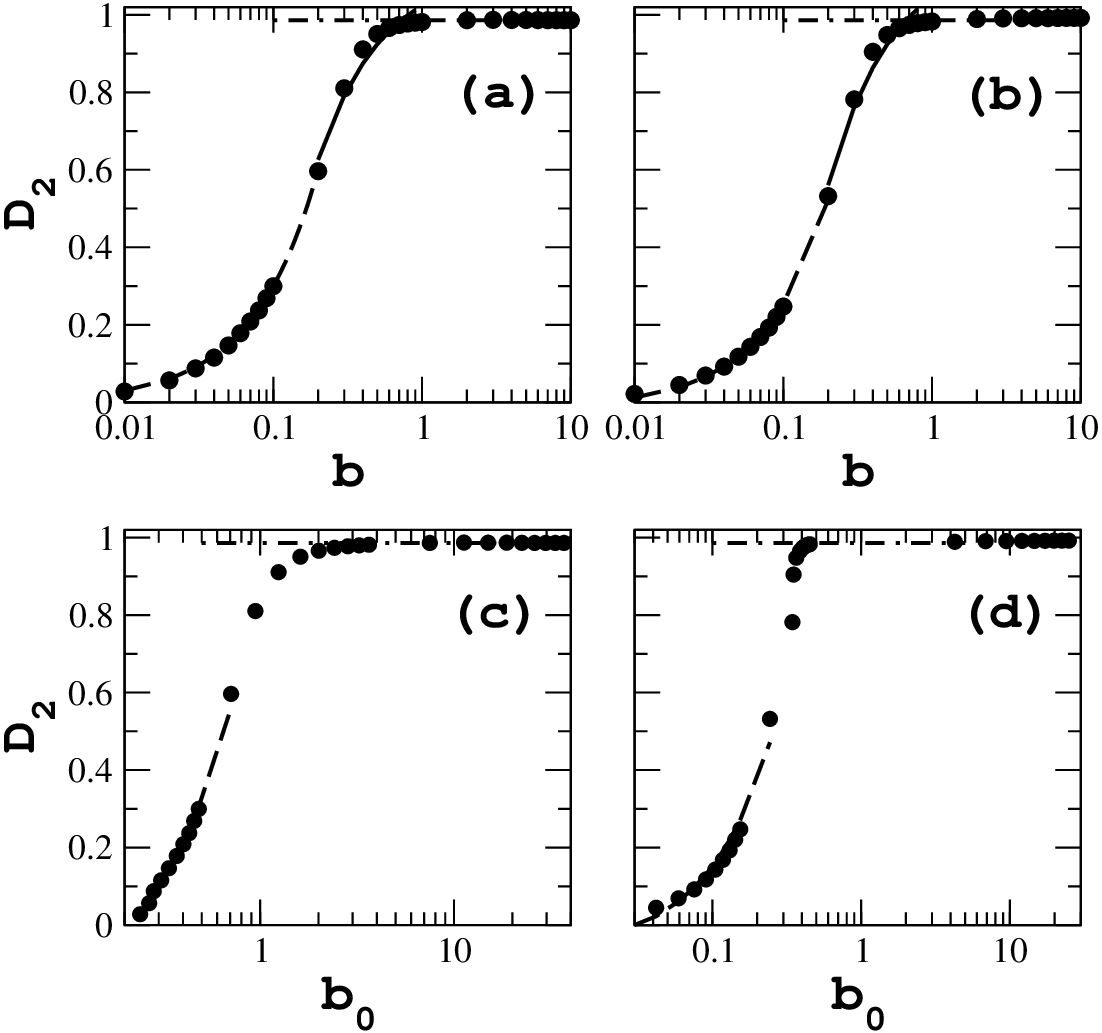}
\caption{(Color online) The fractal dimension $D_2$ of the
critical eigenstates of (a) PRBUM-COE and (b) PRBUM-CUE
as a function of $b$. For the PRBUM-COE case,
in the $b \ll 1$ regime $D_2$ varies linearly with $b$ as $D_2
= 3 b$ (dashed line). In the regime $b \in [0.2, 1.0]$, $D_2$ approximately
follows $D_2 = a - c/b$ (solid line) where the parameters $a \simeq 1.11$ and
$c \simeq 0.10$. Finally, in the $b \gtrsim 1.0$ regime $D_2$
rapidly saturates at a value close to unity (dashed-dot line). For the
PRBUM-CUE case,
$D_2$ behaves qualitatively the same as in the PRBUM-COE case. However,
in the $b \ll 1$ regime, $D_2$ does not exactly follow $D_2 = 3 b$.
In the $b \in [0.2, 1.0]$ regime, $D_2$ again approximately follows
$b$ as $D_2 = a - c/b$ with $a \simeq 1.17$ and $c \simeq 0.12$. The same
$D_2$ is also plotted as a function of $b_0$ for (c) PRBUM-COE and (d) PRBUM-CUE.
There, for the PRBUM-COE case and for $b \in [0.01, 0.2]$ and hence
$b_0\in [0.2, 0.8]$, $D_2 \simeq 1.13 b_0$ (dashed line). In the regime of 
intermediate $b$ values, we cannot fit $D_2$ vs $b_0$ curve by the function 
$D_2 = a - c/b_0$. For even larger $b_0$ values, similar fast saturation of 
$D_2$ is observed. The PRBUM-CUE case behaves qualitatively the same.
}
\label{B_Vs_D2}
\end{figure}

In Fig. \ref{B_Vs_D2}(a) and (b) we show how the fractal dimension $D_2$ of
PRBUM eigenstates for $\alpha = 1.0$ depends on
$b$. For the PRBUM-COE case, in $b \ll 1$ regime, $D_2$ is found to
change linearly with $b$, namely, $D_2
= 3 b $ (dashed line in Fig. \ref{B_Vs_D2}(a)). This feature is
different from the parallel result from the PRBM ensemble at $\alpha
= 1.0$, which gives $D_2 = 2 b$ \cite{mirlin2}. One may wonder if this
difference is simply due to the difference between $b$ and $b_0$. So in Fig.
\ref{B_Vs_D2}(c), we also show how $D_2$ depends on $b_0$. Here we like to
mention that, when $b$ varies from $0.01$ to $0.2$, correspondingly $b_0$
varies from $0.2$ to $0.8$. In this regime, we find $D_2 \simeq 1.13 b_0$
(dashed line in Fig. \ref{B_Vs_D2}(c)), this expression is giving
a proportionality constant far from {\it two}. For the
PRBUM-CUE case, in the same regime, $D_2$ still varies linearly with $b$
(dashed line in Fig. \ref{B_Vs_D2}(b)); but the proportionality constant is
not an integer ($D_2 \simeq 2.68 b$). On the other hand, $D_2$ varies with
$b_0$ as $D_2 \simeq 2.22 b_0$ (dashed line in Fig. \ref{B_Vs_D2}(d)).
By contrast, in the regime $b \in [0.2, 1.0]$, $D_2$ approximately follows
$b$ as $D_2 = a - c/b$ (solid line in Fig. \ref{B_Vs_D2} (a) and (b)) where
$a$ and $c$ are constants. We also find from our
numerical fit that, for the PRBUM-COE case $a \simeq 1.11$ and $c \simeq 0.10$;
whereas for the PRBUM-CUE case, $a \simeq 1.17$ and $c \simeq 0.12$. This
behavior is qualitatively similar to the behavior observed in case of
the PRBM ensemble. For PRBM in the same regime,
$D_2 = 1 - 1/\beta\pi b$ where $\beta = 1$ and $2$ for the orthogonal and
unitary symmetries, respectively \cite{mirlin2}. Note also that
we cannot get a nice fitting with the same form for the $D_2$ vs. $b_0$ curve.
Furthermore, in the regime of
$b \gtrsim 1$, $D_2$ is found to rapidly approach a constant value
$1-\epsilon$, where $\epsilon \simeq 0.013$ for the PRBUM-COE case
and $\epsilon \simeq 0.011$ for the PRBUM-CUE case (dashed-dot line in
Fig. \ref{B_Vs_D2}(a) and (b)). Similar fast saturation of $D_2$ to a
constant value is also observed in $D_2$ vs. $b_0$ curve.
Results in Fig. \ref{B_Vs_D2} also suggest that tuning the $b$ value from
$b=0.1$ to $b=0.5$ as in Fig. \ref{alphVsD2} already covers a significant
range.

\section{Is $\alpha=1.0$ the transition point?}
\label{sec4}

In the previous section, we have shown that the PRBUM
delocalization-localization transition takes place somewhere around $\alpha=1.0$.
The exact transition point is not determined.
Numerically, similar results were also
observed in the case of PRBM. However in the PRBM  case, one can
analytically determine the exact transition
point by mapping the problem onto a nonlinear $\sigma$ model with nonlocal
interactions \cite{prbm-pap}. Because we do not have a good
analytical tool for treating PRBUM, we need to rely on
numerical investigations. So we now further investigate the transition point
from a different perspective.

For PRBM, the `not so sharp transition' was attributed
to the fact that there exists a range of values of $\alpha$ around
{\it unity} in which the eigenstates are fractal up to a characteristic
length denoted by $\xi_{\text{frac}}$. This length is anomalously large as
compared to the system size and it diverges around $\alpha = 1.0$
\cite{anom_large_crit}. Following the procedure given in Ref.
\cite{anom_large_crit},
we now calculate a different kind of fractal dimension $d_2$ of the
eigenstates by employing standard box counting procedure through the
expression
\begin{figure}[t]
\centering
\includegraphics[width=8.4cm,height=4cm]{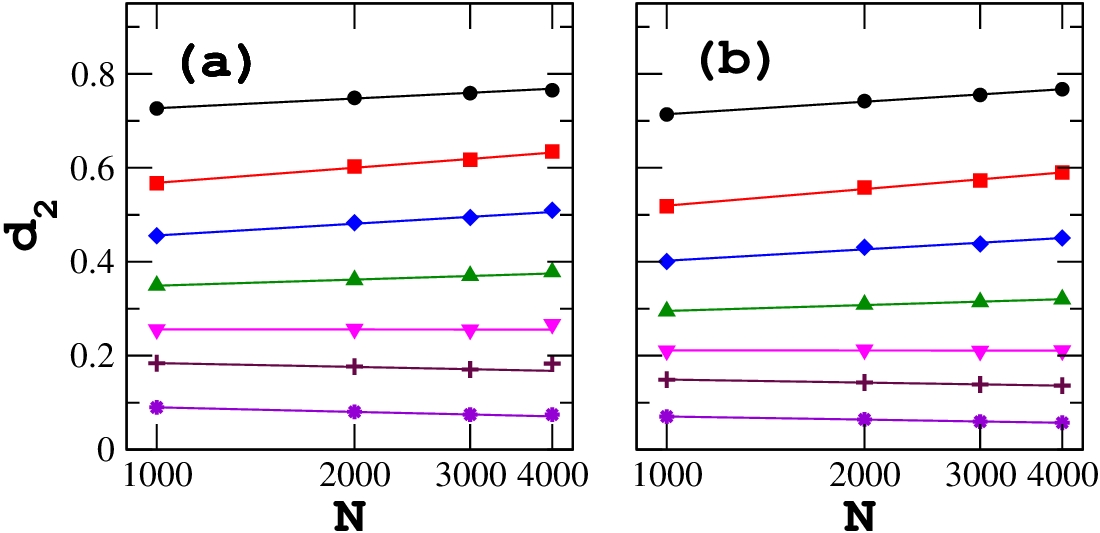}
\caption{(Color online) Ensemble average of the fractal dimension $d_2$ is
calculated using box
counting procedure as a function of the system size $N$ for $\alpha = 0.80,
0.90, 0.95, 1.00, 1.05, 1.10$ and $1.20$ from top to bottom. Here we construct
the ensemble by taking $1200$
to $300$ matrices for different Hilbert space dimensions. 
The results for PRBUM-COE and PRBUM-CUE are presented respectively in panel (a) 
and (b), with $b=0.1$.}
\label{d2vsDim}
\end{figure}

\begin{figure}[b]
\centering
\includegraphics[width=8.4cm,height=4cm]{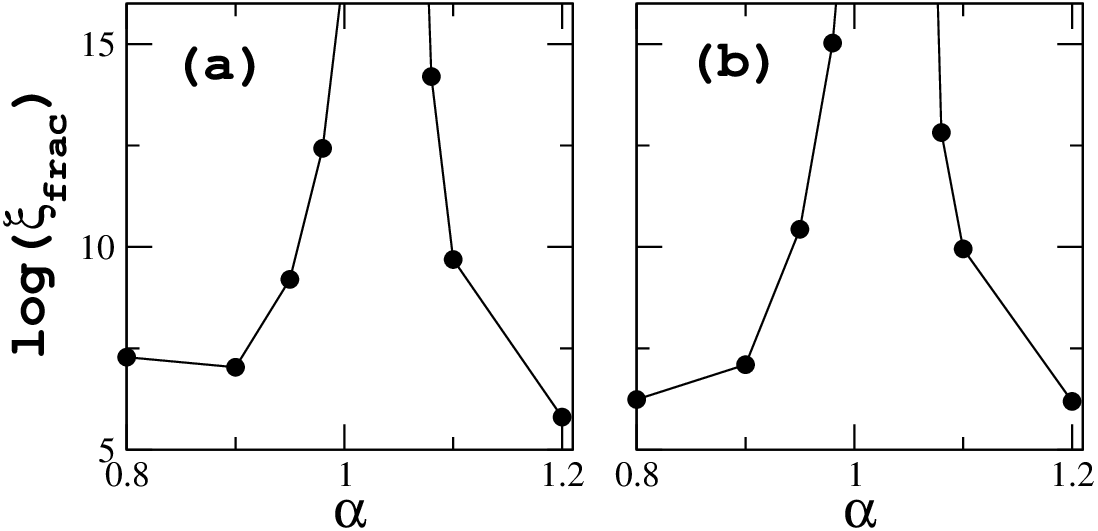}
\caption{Logarithm of the characteristic length
$\xi_{\text{frac}}$ (determined from the results in Fig.~5)
is plotted as a function of $\alpha$ for (a) PRBUM-COE
and (b) PRBUM-CUE. At
$\alpha \approx 1.0$, $\log(\xi_{\text{frac}})$ shows signs of divergence.}
\label{alphVSfrac}
\end{figure}
\be
d_2 \equiv \lim_{\delta \rightarrow 0} \frac{\ln[\chi_2(n)]}{\ln \delta}
\label{calc_d2}
\ee
where $\delta = n/N$ and $\chi_2(n)$ is the $2$nd moment of the probability
density of the eigenstate in the boxes of size $n$. In the delocalized
(metallic) regime, $d_2$ should approach towards $1$
from its fractional value as we increase the system size $N$. So in this
regime, the slope of $d_2$ vs. $N$ curve is expected to be positive. On the
other hand, in the localized regime, the limiting value of $d_2$ should be
$0$, and consequently the slope of the $d_2$ vs. $N$ curve will be negative.
This implies that there exists a critical value of $\alpha$ at which the
slope will be
{\it zero}. This particular value of $\alpha$ can be regarded as a more precise
transition point and on that point $d_2$ will be independent of the
system size (i.e. scale-invariance), at least for the system size we considered.
The characteristic length $\xi_{\text{frac}}$ can be defined
as an extrapolated system size for which $d_2 = 1.0$ and $0.0$ in the delocalized 
and the localized regime, respectively.

To proceed with this new angle we perform ensemble averages of $d_2$ over those 
eigenstates of a matrix that have IPR values close to the maximal IPR
(thus reducing the fluctuations in $d_2$). Here we construct the ensemble by 
taking $1200$ to $300$ matrices depends on the Hilbert space dimension. 
In Fig.~\ref{d2vsDim}, we plot $d_2$ as a function of the system
size $N$ for $\alpha = 0.80, 0.90, 0.95, 1.00, 1.05, 1.10$ and $1.20$ from
top to bottom on a semi-logarithmic scale. The results therein can be regarded
as a type of finite-size scaling analysis (quite different from other approaches,
e.g., \cite{finite-size}). These curves are fitted very
well by straight lines. We then determine $\log(\xi_{\text{frac}})$ as a function
of $\alpha$ from these fitted straight lines.  The results are presented in
Fig.~\ref{alphVSfrac}. For both panels in Fig.~\ref{alphVSfrac}, the left (right) 
part of $\log(\xi_{\text{frac}})$ vs $\alpha$ curve 
shows the fitted results using $d_2=1$ ($d_2=0$).
Similar to the PRBM ensemble case
\cite{anom_large_crit}, we can see that $\log(\xi_{\text{frac}})$ diverges
around $\alpha = 1.0$.

\begin{figure}[t]
\centering
\includegraphics[width=8.4cm,height=4cm]{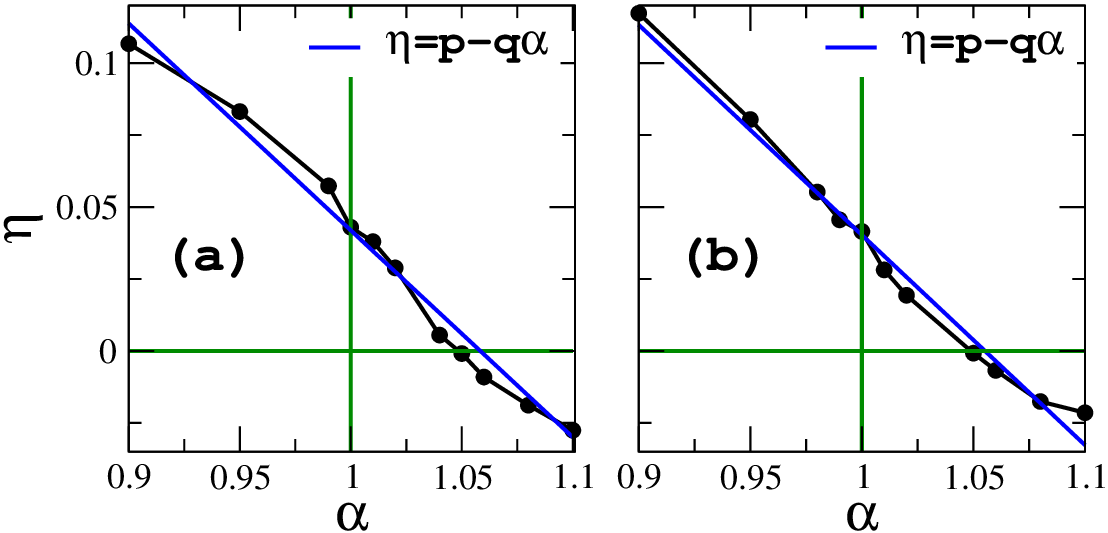}
\caption{(Color online) The slope $\eta$ of $d_2$ vs. $\log(N)$ curves
(as presented in Fig. \ref{d2vsDim}) is plotted as a function of $\alpha$ for
(a) PRBUM-COE and (b) PRBUM-CUE. These results are fitted with a linear function
(solid line) which gives
slope $p = 0.76$ and intercept $q = 0.72$ for the PRBUM-COE case. For the
PRBUM-CUE case, the slope $p=0.73$ and the intercept $q=0.77$. In both panels, the horizontal line represents
$\eta = 0$ and the vertical line indicates $\alpha = 1.0$.}
\label{alphVSslope}
\end{figure}

In Fig. \ref{alphVSslope}, we plot the slope $\eta$ determined from the $d_2$ vs.
$N$ curves in Fig. 5, for different values of $\alpha$. The presented results suggest
that, at $\alpha = 1.0$, $\eta$ is approximately equal to $0.05$.
This slope is small, but not exactly zero. The slope reaches nearest to zero
around $\alpha = 1.05$. In this sense, $\alpha = 1.05$ may be regarded as
a more precise transition point for both PRBUM-COE and PRBUM-CUE cases (applies to
$b=1.0$ as well).  One may suspect that, instead of $\alpha = 1.0$, $\alpha_0 = 1.0$
might be the precise transition point. However, according to Fig. \ref{alphVsalph0},
at $\alpha = 1.05$, the parameter $\alpha_0$ is approximately equal to $0.95$.
So we have to rule out this possibility. Another interesting observation from Fig. 7
is the following: around $\alpha = 1.0$, the slope $\eta$ is essentially
a linear function of $\alpha$. The corresponding linear fitting of
the form $\eta = p - q \alpha$ gives $p \simeq 0.76$ and $q \simeq 0.72$ for
the PRBUM-COE case, and $p \simeq 0.73$ and $q \simeq 0.77$ for the PRBUM-CUE
case.  In both cases, $p\approx q$. If we assume $p = q$ in the thermodynamic
limit, then $\eta \simeq p (1 - \alpha)$, which predicts that
$\eta$ is zero only when $\alpha = 1.0$. This hints that even here $\alpha=1.05$ seems
to be the more precise critical point for $b=0.1$, the slight departure from
$\alpha = 1.0$ can be due to the finite size of our system.
One can also use our PRBUM algorithm to argue in favor of $\alpha = 1.0$ as the true
transition point in the thermodynamics limit. That is,
if $\alpha = 1.0$ is not the transition point, then our algorithm
would have to generate a critical PRBUM ensemble starting from a non-critical PRBM
ensemble.

\section{Statistics of PRBUM eigenvector components}
\label{sec5}

\begin{figure}[t]
\centering
\includegraphics[width=8.4cm,height=6cm]{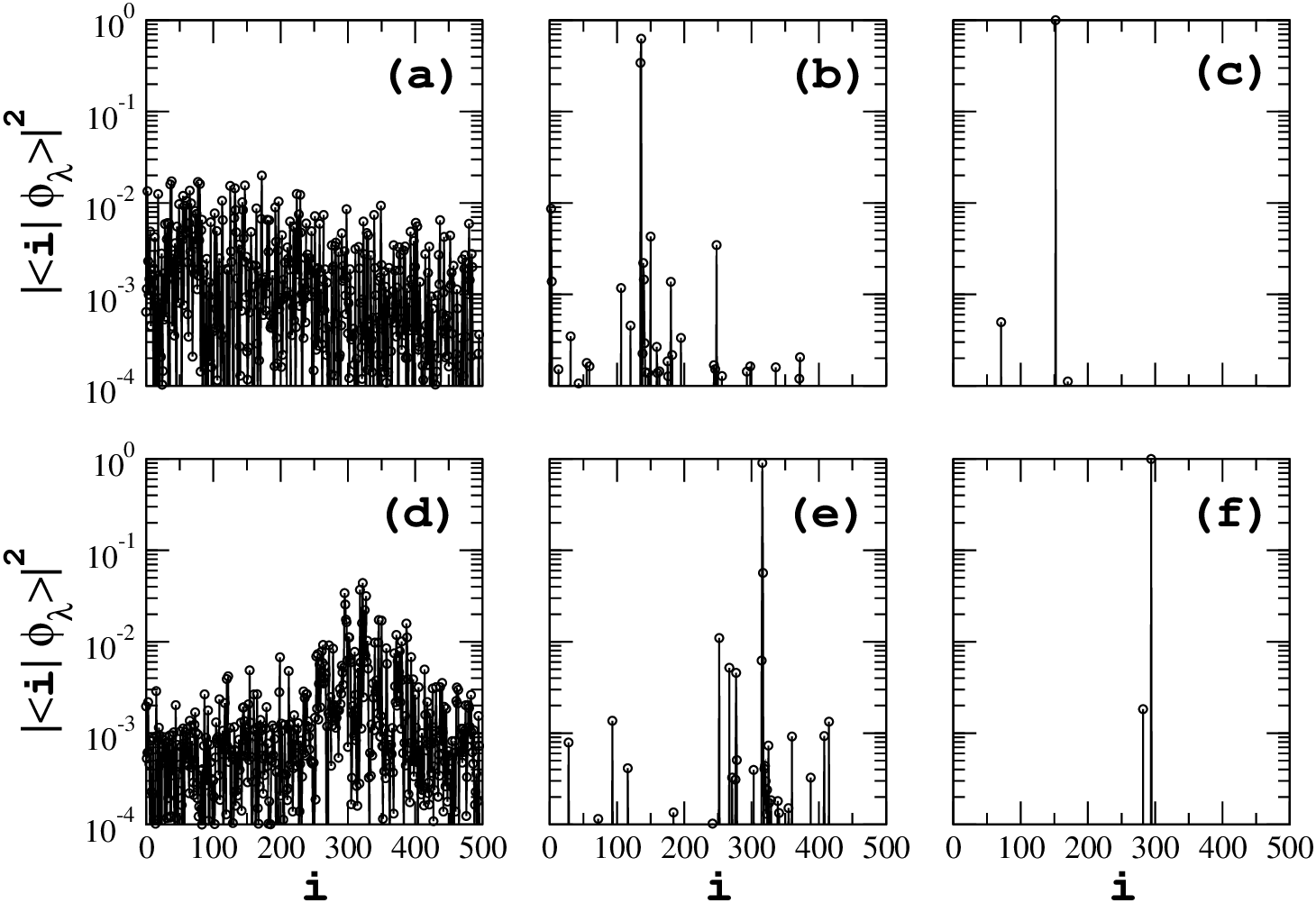}
\caption{(Color online) Typical eigenvector profile for PRBUM-COE
[(a)-(c)] and PRBUM-CUE [(d)-(f)]. In (a) and (d) $\alpha = 0.5$; in
(b) and (e) $\alpha = 1.0$, and in (c) and (f) $\alpha = 1.5$.
$b = 0.1$ is fixed.}
\label{ExmplEvec}
\end{figure}

In this section we investigate the statistical properties of the
eigenvector components of PRBUM. This is interesting because the
parallel results for standard Dyson's circular ensembles provide 
a good reference point for us. Figure~\ref{ExmplEvec} depicts the profile
of some typical eigenvectors for PRBUM-COE and PRBUM-CUE, for three
different  values of $\alpha$, i.e., $\alpha = 0.5, 1.0,$ and $1.5$,
and a fixed $b$ value.  For $\alpha = 0.5$ shown in
Fig.~\ref{ExmplEvec}(a) and  Fig.~\ref{ExmplEvec}(d), both PRBUM-COE
and PRBUM-CUE give delocalized random eigenvectors, which is
consistent with the observation that $D_2\sim 1$ for $\alpha= 0.5$.
Turning to Fig.~\ref{ExmplEvec}(b) and Fig.~\ref{ExmplEvec}(e) for
the case of $\alpha=1.0$, the eigenvectors exhibit a quite
sparse structure, but are not localized in any particular regime.
The same property was observed in the PRBM ensemble
\cite{prbm-pap}.  Finally, for the case of $\alpha = 1.5$ shown in
Fig.~\ref{ExmplEvec}(c) and Fig.~\ref{ExmplEvec}(f), it is seen that
the eigenvectors are highly localized, which is consistent with the
result of $D_2\sim 0$ seen in Fig.~\ref{alphVsD2}.  In the following
we discuss the statistics of the magnitude of PRBUM eigenvector
components.

\begin{figure}[b]
\centering
\includegraphics[width=8.4cm,height=7cm]{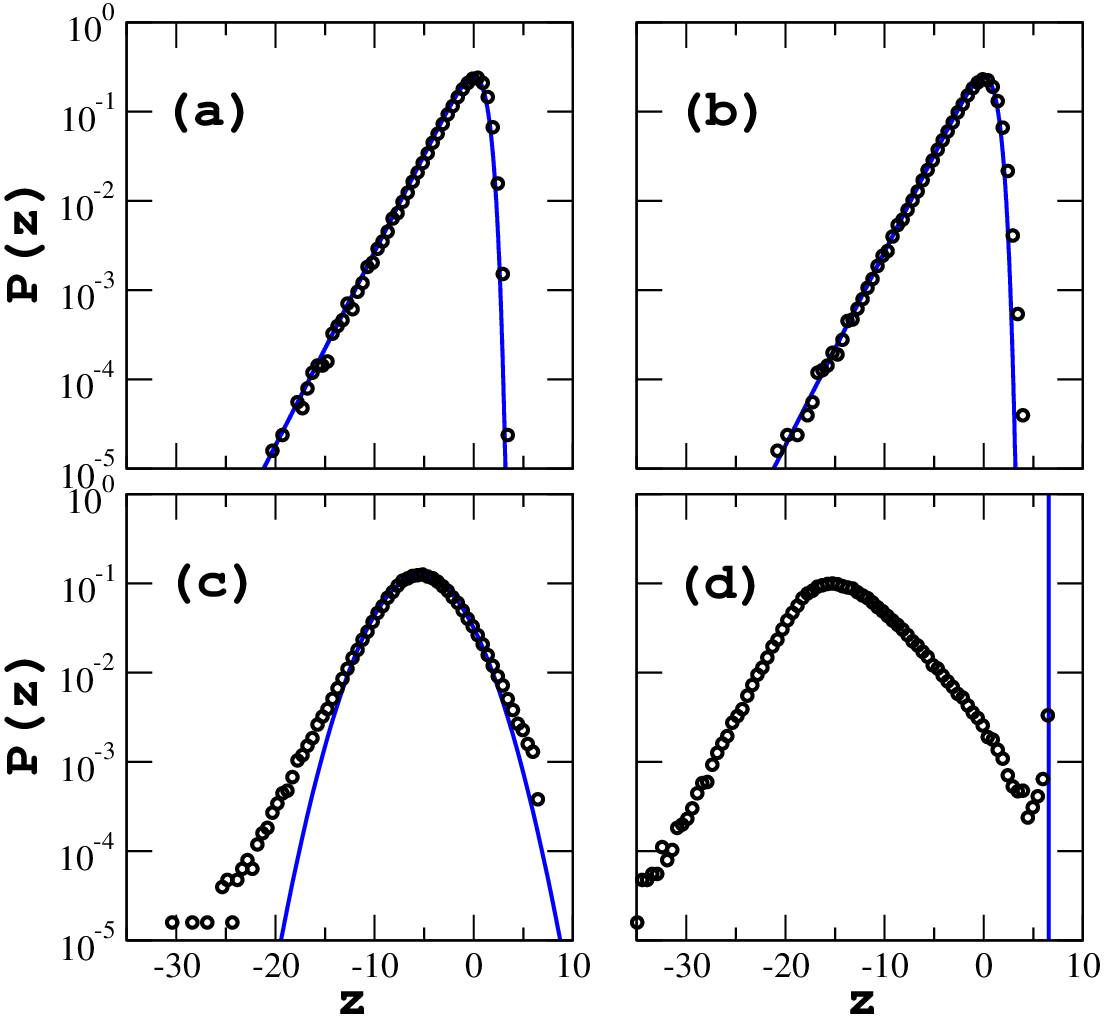}
\caption{(Color online) Distribution function $P(z)$ vs $z$ for the
eigenvector component statistics of PRBUM-COE, where $z = \ln [N
|\langle m|\phi_\lambda\rangle|^2]$ and $N$ is the matrix size. Here
$N = 500$ and $b=0.1$. (a) $\alpha = 0.1$, the solid line represents
the standard COE distribution
of Eq.~(\ref{COE_DistFunc}). (b) Same as in (a) but with $\alpha =
0.5$. (c) $\alpha = 1.0$ (around the critical point). The central part of
$P(z)$ follows closely a normal distribution (solid line) with mean
value $\langle z \rangle \simeq -5.34$, and  variance $\sigma_z
\simeq 3.25$.  Strong deviations from the normal distribution are
seen for $|z| \gg 0$. (d) $\alpha = 1.5$.  An additional peak
(location indicated by the vertical solid line) appears at $z = \ln
N = \ln (500) \simeq 6.22$. This additional peak is due to the
emergence of some most localized states.} \label{COE_EvecDist}
\end{figure}

Figure \ref{COE_EvecDist} shows the distribution function of the
magnitude squared of the eigenvector components for PRBUM-COE,
considering an ensemble of $100$ matrices with $N=500$, for four
different $\alpha$ values but with the $b$ value fixed at $b = 0.1$.
The results are plotted in terms of the probability
distribution function $P(z)$, where $z \equiv \ln y \equiv \ln [N
|\langle m|\phi_\lambda\rangle|^2]$. In the case of $\alpha=0.1$, because
the associated fractal dimensional $D_2\sim 1.0$ (see Fig. \ref{alphVsD2}),
one expects $P(z)$ to follow Dyson's standard COE distribution for
eigenvector components. Because the COE distribution of $y$
is given by~\cite{porter}
\be
p(y) = \frac{1}{\sqrt{2 \pi y}}
\exp\left(-\frac{y}{2}\right),
\ee
the COE distribution for $z$ is given by
\be P_{\text{COE}}(z) =
\frac{1}{\sqrt{2 \pi}} \exp\left[\frac{z - \exp(z)}{2} \right].
\label{COE_DistFunc}
\ee
As seen in Fig.~\ref{COE_EvecDist}(a), the
numerical result (open circle) follows Eq. (\ref{COE_DistFunc})
(solid line) remarkably well. The value of $\alpha$ is then
increased to $\alpha=0.5$ in Fig.~\ref{COE_EvecDist}(b).  It is found
that $P(z)$ is still very close to $P_{\text{COE}}(z)$, with slight
deviation at the right tail.  This is consistent with our early
observation that the $D_2$ value is still close to unity for
$\alpha=0.5$.  The case $\alpha = 1.0$  is shown in
Fig.~\ref{COE_EvecDist}(c), where $P(z)$ is seen to be significantly
different from the COE distribution  $P_{\text{COE}}(z)$. Indeed,
the $P(z)$ for PRBUM-COE for this (at least) near-critical case is much broader than
$P_{\text{COE}}(z)$, reflecting that the possibility of having very
large or very small eigenvector components is
much higher than in the conventional random matrix theory. The solid
line in Fig.~\ref{COE_EvecDist}(c) represents a normal distribution,
and it is seen that the central part of $P(z)$ is close to the normal
distribution. Finally, Fig.~\ref{COE_EvecDist}(d) displays the
result for $\alpha = 1.5$, a value that yields $D_2 \rightarrow 0$
as seen in Fig.~\ref{alphVsD2}.  In this case, many eigenvectors are
highly localized and as a result, one observes a much enhanced
probability at large $z$. Indeed, because the most localized
eigenvector has only one component with magnitude unity and all
other components with magnitude zero, it is now possible to have a
high probability at $z=\ln N$.  The additional peak at $z = \ln 500
\simeq 6.22$ in Fig.~\ref{COE_EvecDist}(d) indicates that some of
the eigenstates are actually close to the most localized states. We
have also checked that if $\alpha$ further increases, then this
additional peak becomes more and more pronounced. At about $\alpha =
10.0$ essentially all the eigenvectors become the most localized
states.

\begin{figure}[t]
\centering
\includegraphics[width=8.4cm,height=7cm]{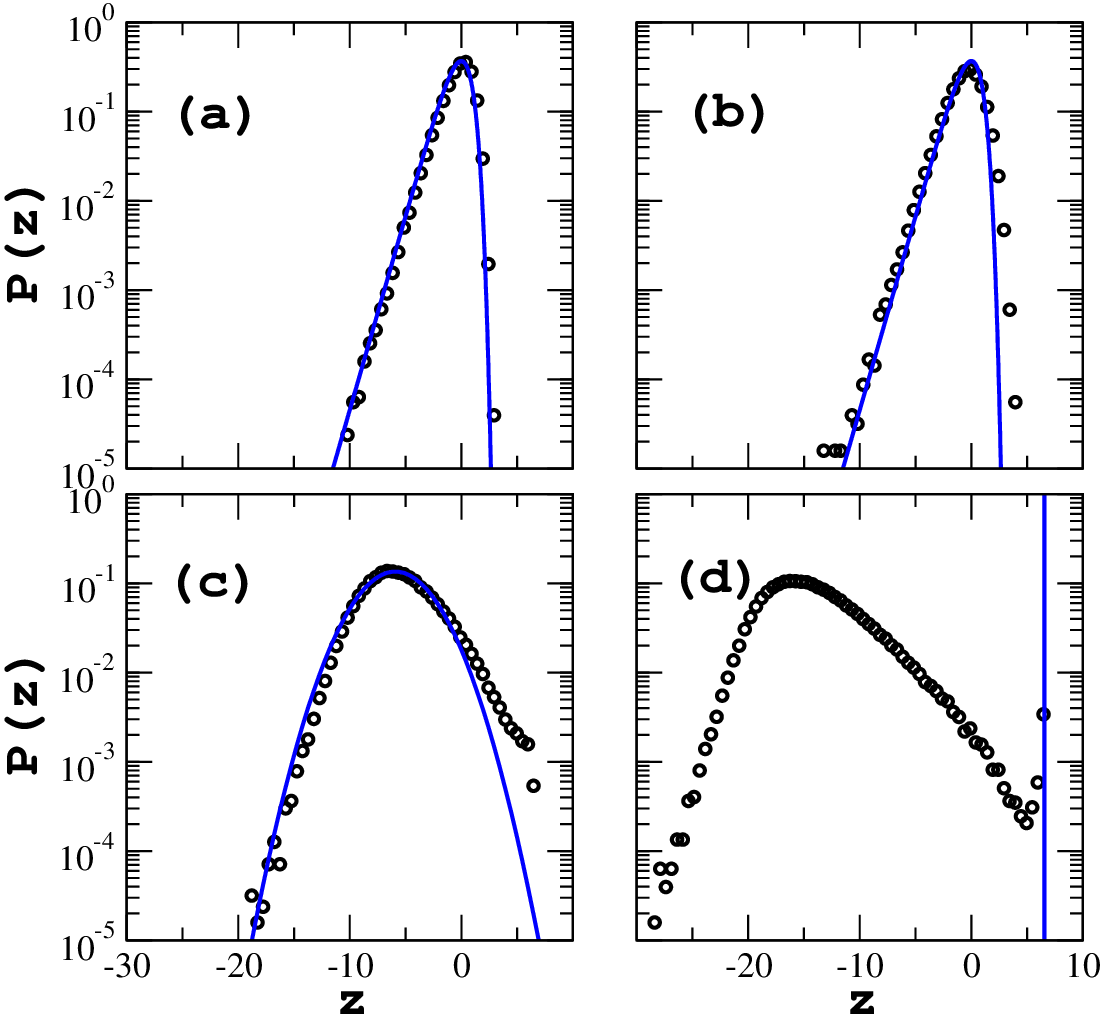}
\caption{(Color online) Same as in Fig.~\ref{COE_EvecDist}, but for
PRBUM-CUE. (a) $\alpha = 0.1$.  The solid line represents the
standard CUE distribution given in Eq. (\ref{CUE_DistFunc}). (b)
Same as (a) but for $\alpha = 0.5$. (c) $\alpha = 1.0$ (around the critical
point).  $P(z)$ follows closely a normal distribution (solid line)
with mean $\langle z \rangle \simeq -5.93$ and  variance $\sigma_z
\simeq 2.95$. Strong deviations from the normal distribution are
seen for $z \gg 0$. The left tail ($P(z)$ at $z \ll 0$) agrees with
the normal distribution better than in the PRBUM-COE case. (d)
$\alpha = 1.5$. An additional peak (location indicated by the
vertical solid line) appears at $z = \ln N = \ln (500) \simeq 6.22$.
This additional peak is due to the emergence of some most localized
states.}
\label{CUE_EvecDist}
\end{figure}

Let us now perform the same analysis for PRBUM-CUE  and the results
are presented in Fig. \ref{CUE_EvecDist}. Here the reference point
is Dyson's CUE statistics of the magnitude squared of eigenvector
components \cite{porter}, i.e., \be p(y) = \exp(-y), \ee which leads
to \be P_{\text{CUE}}(z) = \exp \left[z - \exp(z)\right].
\label{CUE_DistFunc} \ee As seen from Fig.~\ref{CUE_EvecDist}(a),
the numerical result for $\alpha=0.1$ agrees with
$P_{\text{CUE}}(z)$ for the entire range of $z$. This again confirms
that PRBUM essentially reduces to CUE for a sufficient small
$\alpha$.   In the case $\alpha=0.5$ shown in
Fig.~\ref{CUE_EvecDist}(b), $P(z)$ closely follows
$P_{\text{CUE}}(z)$ with marginal deviation in the right tail.  The
case of $\alpha=1.0$ in the delocalization-localization transition regime
is shown in Fig.
\ref{CUE_EvecDist}(c), with features clearly different from that of
$P_{\text{CUE}}(z)$. We have also compared the numerical result with
a normal distribution [solid line in Fig.~\ref{CUE_EvecDist}(c)].
Interestingly, the left part of $P(z)$ as well as its central part
is seen to agree with the fitted normal distribution reasonably
well, but clear deviation from the normal distribution is seen for
$z \gg 0$.  This feature is different from what is observed in the
PRBUM-COE case [see Fig.~\ref{COE_EvecDist}(c)], where deviations
from a normal distribution is also clearly seen in the left tail.
The result for $\alpha = 1.5$ is shown  in Fig.
\ref{CUE_EvecDist}(d). Similar to the PRBUM-COE case in
Fig.~\ref{COE_EvecDist}(d),  here again an additional peak emerges
at $z = \ln(500) \simeq 6.22$, which is clear evidence that some
states become the most localized states.

Summarizing this section, it is observed that when the
delocalization-localization transition occurs, the distribution
function $P(z)$ for PRBUM shows strong deviations from Dyson's
COE/CUE predictions. These deviations may be regarded as one main
characteristic of the Anderson transition. Interestingly,  in the
case of PRBUM-CUE, both the central part and the left tail of $P(z)$
is close to a normal distribution, whereas in the case of PRBUM-COE,
only the central part of $P(z)$ is close to a normal distribution.

\section{Critical spectral behavior of PRBUM}
\label{sec6}

In this section we investigate the manifestations of the
delocalization-localization transition of PRBUM in its NNSD
statistics.  In the Anderson tight-binding model on a $3D$ cubic
lattice, Braun {\it et al}. found that on the metal-insulator
transition point, the NNSD depends on the boundary conditions of the
cubic lattice \cite{braun1}. Upon averaging over different boundary
conditions, it was discovered that the NNSD follows approximately
the semi-Poisson distribution, i.e., \be P_{\text{sp}}(s) = 4 s \exp(-2
s), \label{SP1} \ee where $s\ge 0$ is the spacing between two
neighboring levels. In the $s \rightarrow 0$ limit,
$P_{\text{sp}}(s) \sim 4 s$, which is a linear level repulsion
property shared by Dyson's COE statistics and many time-reversal
symmetric dynamical systems with chaotic classical limits
\cite{mehta,haakebook}.  However, the proportionality constant in
the linear level repulsion for $P_{\text{sp}}(s) \sim 4 s$ is
different from Dyson's COE statistics $P_{\text{COE}}(s) \sim
\frac{\pi}{2} s$. In the other limit $s \gg 0$, $P_{\text{sp}}(s)$
decays exponentially, i.e., $P_{\text{sp}}(s) \sim \exp(-2 s)$,
which is reminiscent of an uncorrelated energy spectrum of a regular
quantum system whose NNSD follows the Poisson distribution
$P_{\text{p}}(s) = \exp(-s)$. Hence in this second limit
$P_{\text{sp}}(s)$ differs significantly from $P_{\text{COE}}(s)$
which decays to zero as $\exp(-\pi s^2/4)$. Therefore, the
semi-Poisson distribution $P_{\text{sp}}(s)$ can be considered as an
intermediate critical NNSD behavior. In Ref. \cite{braun2}, the NNSD
of the PRBM ensemble was also studied. The finding is that the
central part of the NNSD follows $P_{\text{sp}}(s)$ well, but
non-negligible deviations from $P_{\text{sp}}(s)$ were observed for
$s \rightarrow 0$ or for $s \gg 0$ regime.  We are thus motivated to
carry out similar NNSD studies for PRBUM in the Anderson transition
regime and also use the semi-Poisson distribution
as one of our diagnostic tools.  Note also that, since Fig.
\ref{qe-density} already confirms the uniform distribution of PRBUM
eigenphases, we only need to introduce a simple re-scaling factor
$N/2\pi$ to set the average level spacing at unity.

\begin{figure}[t]
\centering
\includegraphics[width=8.4cm,height=7cm]{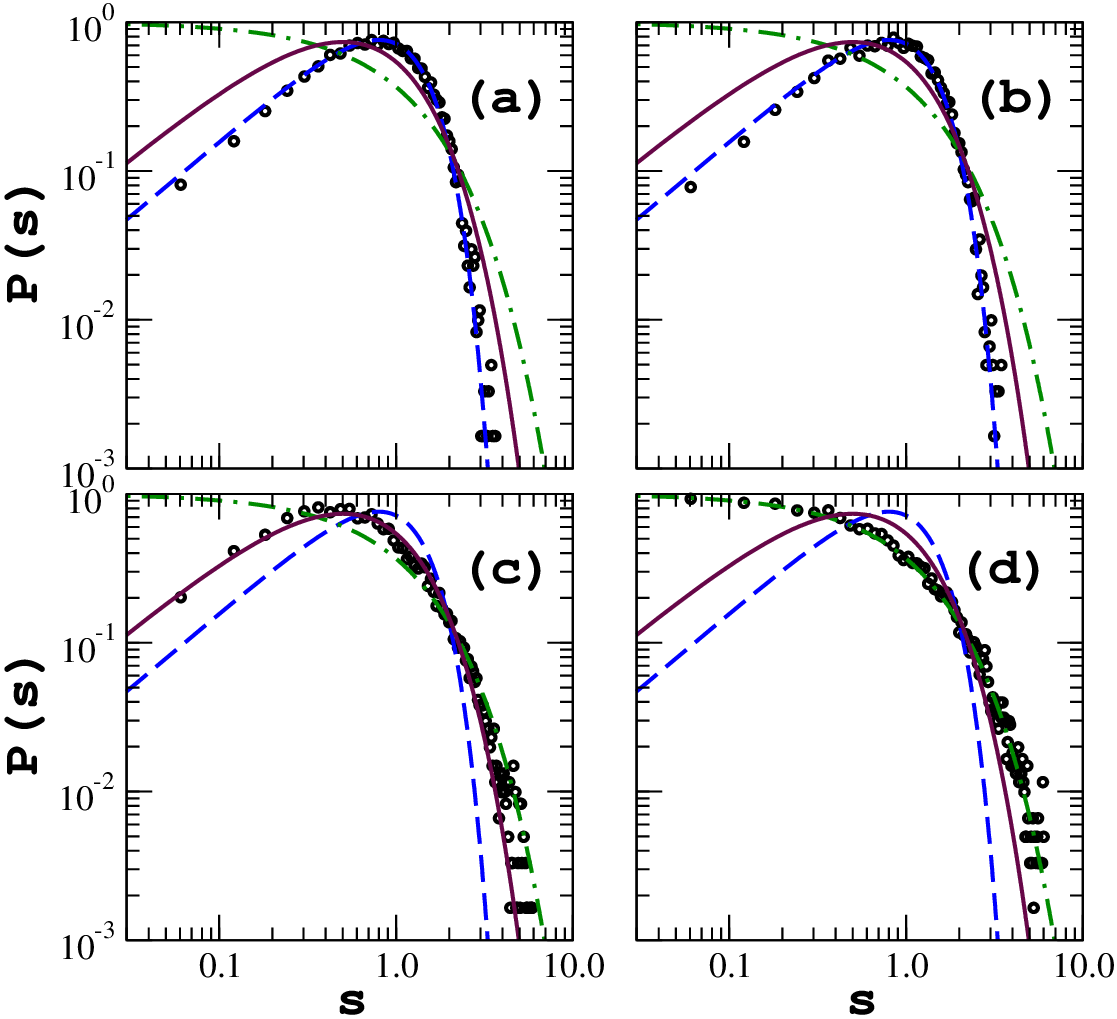}
\caption{(Color online) The NNSD [denoted $P(s)$] of the eigenphases
of PRBUM-COE (open circles) vs level spacing $s$,  for four different
values of $\alpha$. For the sake of comparison, the Wigner-Dyson
distribution for COE (dashed line), the semi-Poisson distribution as
given in Eq.(\ref{SP1}) (solid line), and the Poisson distribution
(dashed-dotted line) are all plotted together. (a) $\alpha = 0.1$.
(b) $\alpha = 0.5$. (c) $\alpha = 1.0$. (d) $\alpha = 1.5$. An ensemble
of 100 matrices with $N=1000$ is used for statistics. Results here
should be connected with those in Fig.~\ref{COE_EvecDist}.}
\label{COE_SpcDist}
\end{figure}

For the same four PRBUM-COE cases as in Fig.~\ref{COE_EvecDist},
Fig.~\ref{COE_SpcDist} depicts the associated NNSD. In case of
$\alpha = 0.1$ in Fig.~\ref{COE_SpcDist}(a), the numerically
obtained NNSD (denoted by $P(s)$ and represented by open circles) is in
good agreement with the Wigner-Dyson statistics $P_{\text{COE}}(s)$
from the conventional random matrix theory (dashed line). For the
sake of comparison, in all the four figure panels we have plotted
the semi-Poisson distribution $P_{\text{sp}}(s)$ (solid line) as
given in Eq. (\ref{SP1}) as well as the Poisson distribution
$P_{p}(s)$ (dashed-dotted line). The case of $\alpha=0.5$ in Fig.
\ref{COE_SpcDist}(b) is similar.  Consider next the critical case in
Fig. \ref{COE_SpcDist}(c). Quite unexpectedly, $P(s)$ follows the
semi-Poisson distribution $P_{\text{sp}}(s)$ very well in the
small-$s$ regime. This agreement in the small-$s$ regime is absent
in PRBM on the critical point. In the large-$s$ regime, deviations
from the semi-Poisson distribution can be seen clearly in
Fig.~\ref{COE_SpcDist}(c). Indeed, $P(s)$ in the large-$s$ regime is
seen to lie between the semi-Poisson distribution and the Poisson
distribution. This feature is also different from one case study of
the $P(s)$ of PRBM \cite{braun2}. In particular, in Ref.
\cite{braun2} the $P(s)$ for PRBM actually lies between
$P_{\text{COE}}(s)$ and $P_{\text{sp}}(s)$, instead of between
$P_{\text{sp}}(s)$ and $P_\text{p}(s)$. This difference can be
explained as follows. For $b = 1.0$ considered in Ref. \cite{braun2},
the fractal dimension of PRBM eigenstates is $D_2 \simeq 0.66$;
whereas here $b=0.1$ and $D_2 \sim 0.28$. As such, the $D_2$ value
in Ref. \cite{braun2} is closer to unity and the NNSD result is
expected to be closer to the conventional random matrix theory.
Furthermore, the much smaller $D_2$ value for our PRBUM considered
here should imply more localized states, and hence the $P(s)$ is
closer to the Poisson distribution. Consistent with this
explanation, the case of $\alpha=1.5$ in Fig.~8(d)
has $D_2\sim 0$, and the associated $P(s)$ indeed agrees well with
the Poisson distribution (dashed-dot line).

\begin{figure}[t]
\centering
\includegraphics[width=8.4cm,height=7cm]{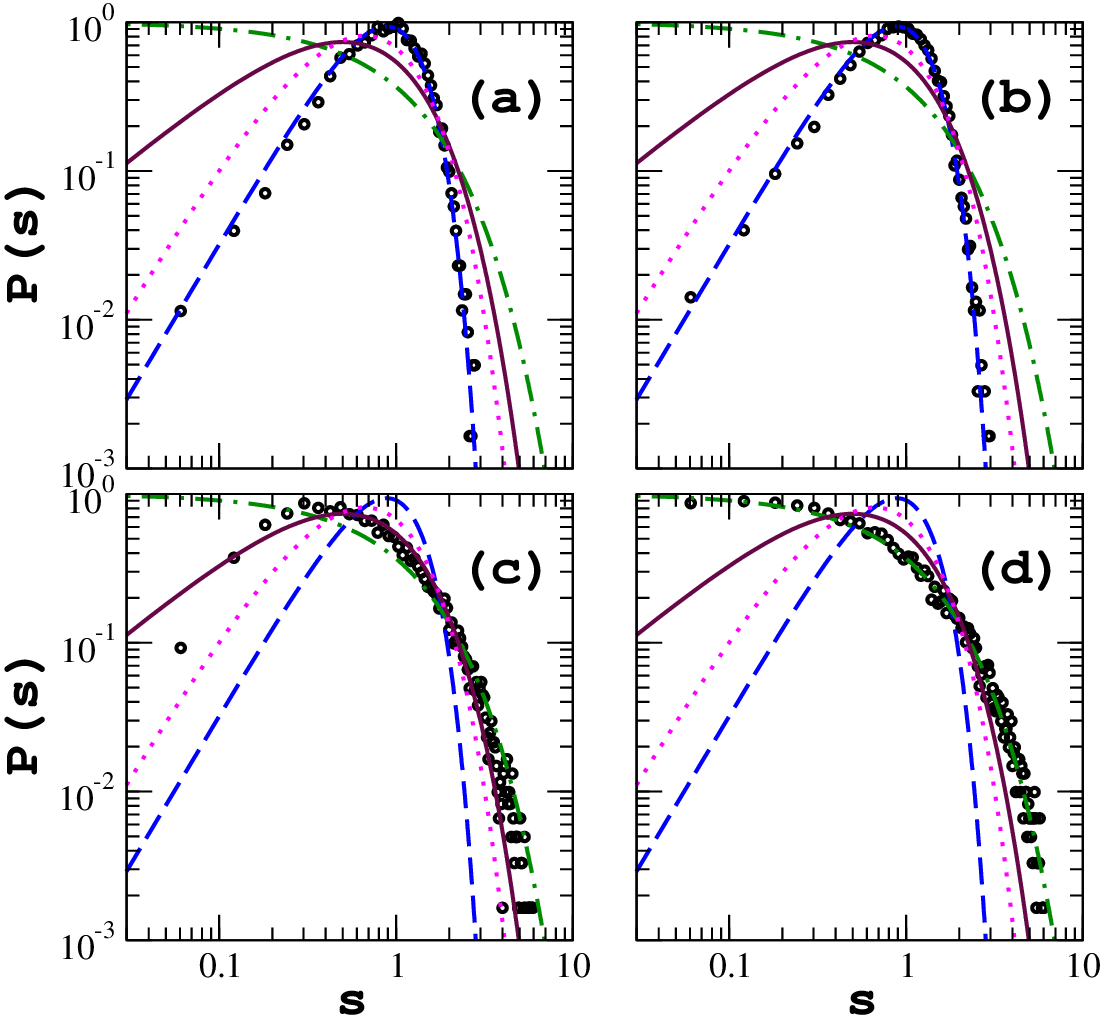}
\caption{(Color online) Same as in Fig.~\ref{COE_SpcDist}, but for
PRBUM-CUE. Here the dashed line represents the Wigner-Dyson
distribution for CUE, and the dotted line represents the generalized
semi-Poisson distribution given in Eq. (\ref{GSP}) with $\beta = 2.0$.
The dashed-dotted line is still for the Poisson distribution and
and the solid line is still for the semi-Poisson distribution.
An ensemble of 100 matrices with $N=1000$ is used for statistics.
Results here should be compared with those in Fig.~\ref{CUE_EvecDist}.}
\label{CUE_SpcDist}
\end{figure}

We next examine the NNSD for PRBUM-CUE in Fig.~\ref{CUE_SpcDist}.
The standard Wigner-Dyson distribution for CUE with quadratic level
repulsion is now plotted as dashed lines in all the four panels. As
expected, $P(s)$ in the cases of $\alpha=0.1$ and $\alpha=0.5$ in
Fig.~\ref{CUE_SpcDist}(a) and Fig.~\ref{CUE_SpcDist}(b) agree with
the standard result from random matrix theory. Similarly, Fig.
\ref{CUE_SpcDist}(d) shows that $P(s)$ for PRBUM-CUE follows the
Poisson distribution for $\alpha = 1.5$, indicating that this
$\alpha$ value is far beyond the delocalization-localization
transition point.  Much more interesting is the case of
$\alpha=1.0$.  Similar to the PRBUM-COE case, in the large-$s$
regime $P(s)$ is found to lie between the semi-Poisson distribution
and the Poisson distribution. For the small-$s$ regime, since $P(s)$
in the PRBUM-COE case nicely follows a semi-Poisson distribution,
one might naively expect that here, due to the breaking of
time-reversal symmetry, the $P(s)$ should follow a generalized
semi-Poisson distribution \cite{bogomolny} with quadratic level
repulsion.  To check if this expectation is correct, in all the
panels of Fig. \ref{CUE_SpcDist} we have plotted an additional curve
(dotted line) to represent a {\it generalized} semi-Poisson
distribution $P_{\text{sp};\beta}(s)$ with $\beta=2$, i.e., \be
P_{\text{sp};\beta}(s) = A_\beta s^\beta \exp[-(\beta+1)
s],~~A_\beta = \frac{(\beta + 1)^{\beta+1}}{\Gamma(\beta+1)}.
\label{GSP} \ee Note that
$P_{\text{sp};\beta=1}(s)=P_{\text{sp}}(s)$.  As seen in Fig.
\ref{CUE_SpcDist}(c), in the small-$s$ regime, the $P(s)$ for
PRBUM-CUE deviates from both $P_{\text{sp};\beta}(s)$ with $\beta=2$
and the semi-Poisson distribution $P_{\text{sp}}(s)$. This implies
the possibility of anomalous level repulsion, namely, $P(s) \sim
s^\beta$ with $\beta$ being a non-integer.

\begin{figure}[t]
\centering
\includegraphics[width=8.4cm,height=7cm]{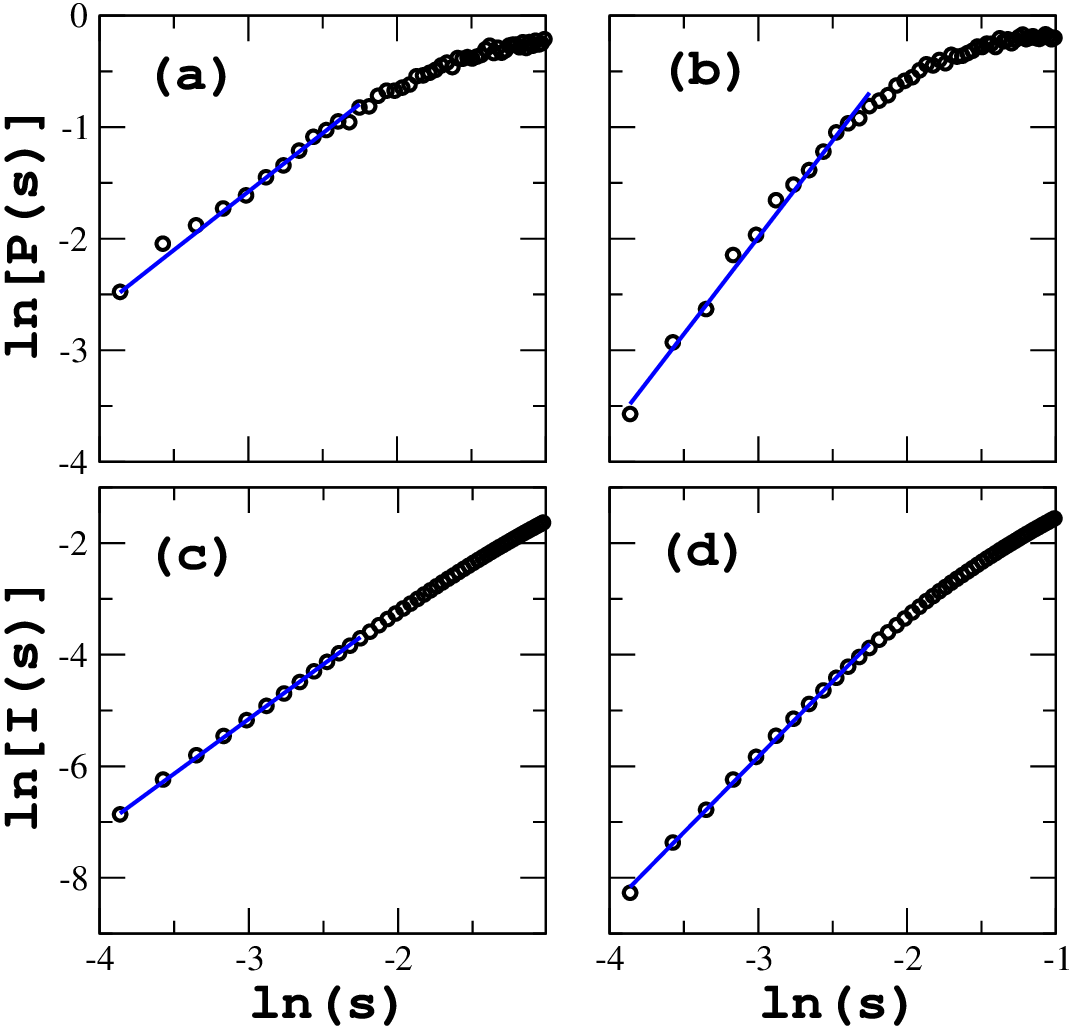}
\caption{(Color online) The NNSD [denoted $P(s)$] and the integrated
nearest neighbor spacing statistics [denoted $I(s)$]  in the
small-$s$ regime for both PRBUM-COE and PRBUM-CUE.  In the top two
panels, the numerical data (open circles) are fitted by $P(s)\sim
s^\beta$.  In the case of PRBUM-COE, $\beta \simeq 1.04$ (almost
linear repulsion), whereas in the case of PRBUM-CUE, $\beta \simeq
1.73$ (signature of anomalous level repulsion). For the bottom two
panels, $I(s)$ vs $s$ is fitted by $I(s)\sim s^{\beta^\prime}$, with
$\beta^\prime \simeq 2.01$ for PRBUM-COE and $\beta^\prime \simeq 2.71$
for PRBUM-CUE, thereby further confirming the found $\beta$ values. To
improve the statistics in the small-$s$ regime, an ensemble of 1000
matrices with $N=1000$ is used for our statistics.}
\label{SpcDist_Small_S}
\end{figure}

In order to investigate carefully the anomalous level repulsion for
critical PRBUM-CUE in the small-$s$ regime, we show in
Fig.~\ref{SpcDist_Small_S} the $\ln[P(s)]$ vs. $\ln(s)$ plot in the
small-$s$ regime ($s<0.36$) and then fit it using the function
$P(s)\sim s^{\beta}$.  To show the marked difference between
PRBUM-CUE and PRBUM-COE around the critical point, this fitting is also
performed for PRBUM-COE.  We obtain $\beta\simeq 1.04$ for
PRBUM-COE, which confirms the linear level repulsion observed
earlier in Fig.~\ref{COE_SpcDist}. By contrast, for PRBUM-CUE, we
find $\beta \simeq 1.73$, clear evidence of anomalous level
repulsion.  To improve the statistics in the small-$s$ regime, we
increased the ensemble size by a factor of ten, i.e., now an ensemble
of 1000 matrices with $N=1000$ is used for our statistics. We have
further checked that the change in the $\beta$ values thus found
numerically is negligible if we further increase the ensemble size
by a factor of three.  Thus, at least for the small-$s$ regime we
considered, the level repulsion is found to be anomalous.  In the
bottom two panels of Fig.~\ref{SpcDist_Small_S}, the numerical
fitting of the integrated nearest-neighbor spacing distribution,
denoted by $I(s)$, is also presented. The fitting curve $I(s)\sim
s^{\beta^\prime}$ yields $\beta^\prime \simeq 2.01$ for PRBUM-COE and
$\beta^\prime \simeq 2.71$ for PRBUM-CUE [Note that, by definition of
$I(s)$, $\beta^\prime = \beta + 1$]. This further supports our
earlier observation that $P(s)$ displays an anomalous level
repulsion in the case of PRBUM-CUE.

\section{Concluding remarks}
\label{sec7}

The Hermitian PRBM ensemble has been studied extensively in the
literature as a standard model for the Anderson transition.  In terms
of the power-law decay of the magnitude of the matrix elements from
the diagonal, the unitary PRBUM ensemble is similar to PRBM
(the generation of PRBUM is much more involved).  This work showed
that, statistics of PRBUM in the
delocalization-localization transition regime can be different from
their PRBM counterparts. Hence PRBUM is not a trivial unitary
analog of the Hermitian PRBM ensemble.
Somewhat consistent with this conclusion, we have observed that for a given
set of $b$ and $\alpha$, the matrix elements of PRBUM as generated from our
algorithm decay from the diagonal as a different power-law function
[see Eq.~(\ref{var-PRBUM}) and Fig.~1, where $b_0$ and $\alpha_0$ can be appreciably
different from $b$ and $\alpha$].  Given our detailed findings, we hope that
PRBUM can capture some critical behavior of actual Floquet eigenstates in the
delocalization-localization transition regime, by modeling system's periodic unitary evolution
 matrix with
power-law decaying matrix elements. Different choices of ensemble parameters
can model a system in either insulating or metallic phase, and if the ensemble parameters
are close to  critical values, then fractal behavior of the Floquet eigenstates (within a certain scale)
may be also modeled by PRBUM.  One obvious advantage of PRBUM as compared
with PRBM is that in performing the statistics all the eigenstates are treated
under equal footing, rather than selecting certain energy windows
as often done in PRBM studies.  We also note that in
our early study \cite{JJJ} PRBUM can correctly model how the
variance of $D_2$ of Floquet eigenstates (in a double-kicked top model)
scales with the system size but PRBM cannot.

In summary, we have numerically studied the statistical properties
of PRBUM in the delocalization-localization regime and have compared
them with PRBM and the standard random matrix theory when feasible.
The dependence of the fractal dimension of PRBUM eigenvectors upon ensemble
parameters is used to examine where the delocalization-localization transition point is, within
some uncertainty due to finite-size effects. For cases
very close to the  transition point,
the central part of the PRBUM eigenvector-component distribution function is
close to a log-normal distribution and the tails (especially the
PRBUM-COE case) show strong deviations from the log-normal distribution.
We have also examined the NNSD of PRBUM  and have
compared the result with the semi-Poission distribution and a
generalized semi-Poisson distribution.
An anomalous level repulsion of PRBUM in the absence of time-reversal symmetry is also
numerically found.

This work was supported by the Academic Research Fund Tier I,
Ministry of Education, Singapore (grant No. R-144-000-276-112).

\end{document}